\documentstyle[12pt,epsf]{ioplppt}
\def\uu{^{\mbox{ }}}
\def\s{\sigma}
\def\bs{\vec{\s}}
\def\u{\uparrow}
\def\up{\u}
\def\dd{\downarrow}
\def\down{\dd}
\def\en{\epsilon}
\def\las{\langle}
\def\ras{\rangle}
\def\la{\left\las}
\def\lla{\la\la}
\def\llas{\las\las}
\def\ra{\right\ras}
\def\rra{\ra\ra}
\def\rras{\ras\ras}
\def\k{\vec{k}}
\def\R{\vec{R}}
\def\S{\vec{S}}
\def\th{$^{\rm th}$\ }
\def\ec{e}

\begin{document}\jl{3}

\title{Electronic structure and resistivity of the\\ double exchange
model}[Electronic structure and resistivity of the double exchange model]

\author{DM Edwards\dag, ACM Green\dag\ and K Kubo\ddag}

\address{\dag\ Department of Mathematics, Imperial College, London SW7 2BZ,
UK}

\address{\ddag\ Institute of Physics, University of Tsukuba, Tsukuba,
Ibaraki 305--8571, Japan}

\begin{abstract}
The double exchange (DE) model with quantum local spins $S$ is studied; an
equation of motion approach is used and decoupling approximations analogous
to Hubbard's are made. Our approximate one-electron Green function $G$ is
exact in the atomic limit of zero bandwidth for all $S$ and band filling
$n$, and as $n\rightarrow 0$ reduces to a dynamical coherent
potential approximation (CPA) due to Kubo; we regard our approximation as
a many-body generalisation of Kubo's CPA. $G$ is calculated
self-consistently for general $S$ in the paramagnetic state and for
$S=1/2$ in a state of arbitrary magnetization. The electronic structure is
investigated and four bands per spin are obtained centred on the atomic
limit peaks of the spectral function. A resistivity formula appropriate to 
the model is derived from the Kubo formula and the paramagnetic state
resistivity $\rho$ is calculated; insulating states are correctly obtained at 
$n=0$ and $n=1$ for strong Hund coupling. Our prediction for $\rho$ is
much too small to be consistent with experiments on manganites
so we agree with Millis \etal
that the bare DE model is inadequate. We show that the agreement with
experiment obtained by Furukawa is due to his use of an unphysical
density of states.
\end{abstract}


\section{Introduction}\label{sIntroduction}

Manganite compounds exhibiting colossal magnetoresistance (CMR) are
of the form
La$_{1-x}$D$_x$MnO$_3$ with D divalent, e.g.\ Ca, Sr, Ba. As the doping
$x$ and temperature $T$ are varied a rich variety of phases are observed,
as discussed by Ramirez \cite{rRamirez}. Recently there has been a lot of
interest in these compounds with $x\stackrel{\tiny >}{\tiny _\sim}0.15$
owing to their interesting magnetotransport properties in this regime: as
$T$ is decreased they undergo a transition to ferromagnetic order, and
near the Curie temperature $T_{\rm C}$ the $T$-dependence of the
resistivity $\rho$ changes from insulating ($\partial\rho/\partial T<0$
for $T>T_{\rm C}$) to metallic ($\partial\rho/\partial T>0$ for
$T<T_{\rm C}$), with a strong peak in $\rho$ at the crossover. The
application of a strong ($\sim 5$T) magnetic field substantially reduces
this peak in $\rho$ and shifts it to higher temperature, giving rise to a
very large negative magnetoresistance. The physical processes causing this 
behaviour have been the subject of much discussion.

The simplest model proposed for the CMR compounds--- the one that we will
study in this paper--- is Zener's \cite{rZener} double exchange (DE) model
with Hamiltonian
\begin{equation}\label{eH}
H=\sum_{ij\s}t_{ij}c^{\dag}_{i\s}c_{j\s}\uu-J\sum_i{\S}_i \cdot\bs_i
=H_0+H_1.
\end{equation}
Here $i$ and $j$ refer to sites of the (approximately) simple cubic lattice
of Mn atoms, $c_{j\s}\uu$ ($c^{\dag}_{i\s}$) is a $\s$-spin conduction
electron annihilation (creation) operator, ${\S}_i$ is a local spin
operator, $\bs_i$ is a conduction electron spin operator,
$t_{ij}$ is the hopping integral with discrete Fourier transform $t_{\k}$,
and $J>0$ is the Hund's rule coupling constant.
The number of conduction electrons per atom $n$
is assumed to be given by $n=1-x$. Physically, the
relevant electrons are those coming from the Mn atoms' 3d shells, which
contain four electrons per site in the undoped compounds and are split by
the cubic crystal field into triply degenerate $t_{2g}$ levels and higher
energy doubly degenerate $e_g$ levels. Strong Hund coupling attempts to
align all electron spins on a site, so the $t_{2g}$ electrons are treated
as localized $S=3/2$ spins while the conduction band is formed from the
$e_g$ states. The main physical effects neglected by $H$ are the double 
degeneracy of the $e_g$ conduction band, impurity atom (D) disorder
scattering, and coupling to the lattice degrees of freedom.

Furukawa \cite{rFurukawa} has studied the infinite dimensional limit of $H$
for $S=\infty$ using dynamical mean field theory, and has concluded that
the DE model's predictions for $\rho$ in the paramagnetic state are
compatible with experiment. Millis \etal \cite{rMillis,rMillis2}
however have claimed that $\rho$ predicted by the DE model is much smaller
than that measured, and that to get agreement with experiment dynamical
Jahn-Teller phonon coupling must be included in $H$. Experiments show that 
phonon coupling is important, with for example a large shift in $T_{\rm C}$
observed upon replacing some of the O atoms with a different isotope
\cite{rZhao}, so Furukawa's claim is puzzling. In this paper we will study 
the one-electron local Green function $G$ and calculate $\rho$ in an
attempt to reconcile the results of Millis and Furukawa. This work has
been briefly summarised elsewhere \cite{rUs,rThey} and is discussed in more
detail by Green \cite{rMe}.

Our starting point is Kubo's calculation based on a dynamical coherent
potential approximation (CPA).
For finite local spins $S$ dynamical scattering processes may occur in which
local spins and conduction electrons exchange angular momentum,
whereas in the classical $S\rightarrow\infty$ limit taken by most authors
the local spins are rigid and $H$ is a one-electron Hamiltonian with
spin dependent
diagonal disorder. Kubo's CPA is an extension of the familiar alloy
CPA which takes these dynamical processes into account in a local
approximation. Since it is a one-electron theory Kubo's approximation is only
valid in the low-density $n\rightarrow 0$ limit, but in this limit
the behaviour of the spectral function is qualitatively correct with bands 
with the correct weights forming about the two atomic limit ($t_{ij}
\rightarrow 0$) peaks as the hopping $t_{ij}$
is switched on. If $J\gg t_{ij}$
double occupation of a site is forbidden so that
at half-filling ($n=1$) the system should be a Mott insulator. This is not the
case in Kubo's CPA where the Fermi level lies within the lower band.
The correct behaviour will be obtained in an approximation which becomes
exact in the atomic limit for all
filling, so we are looking for a many-body
extension of Kubo's CPA, valid for all $n$, which reduces to
Kubo's CPA as $n\rightarrow 0$ and to the correct
atomic limit as $t_{ij}\rightarrow 0$, for all $n$.

It is difficult to extend the usual CPA method to the many-body case.
Instead we return to the original approach of Hubbard \cite{rHubbard3}
in which he applied the equation of motion method to calculate the
one-electron local Green function $G$ for the Hubbard model. His decoupling
approximation was motivated by the alloy analogy in which electrons of one
spin are considered as frozen on atomic sites. Hubbard's `scattering
correction' is equivalent to a CPA treatment of the alloy
analogy \cite{rFukuyama?} and his `resonance broadening correction' was an
attempt to restore some dynamics to the frozen electrons. The idea of
$\downarrow$ spins being frozen in the calculation of the $\uparrow$ spin
Green function $G_{\up}$ is introduced in the equation of motion method by
neglecting commutators of the kinetic part of the Hamiltonian $H_0$ with
$\down$-spin occupation numbers.

In the derivation of our approximation to $G$ for the DE model we will
make approximations analogous to those used by Hubbard to obtain the
scattering correction; terms corresponding to his resonance broadening
correction will be neglected. Our method represents a considerable
extension of Hubbard's owing to the more complicated form of the
interaction term of the DE model, which for instance allows electrons to
change spin via exchange of angular momentum with the local spins. This
effect couples the equations for $G_{\u}$ and $G_{\dd}$ which may perhaps be
regarded as including some resonance broadening effects.
Our choice of
approximations will be guided by the requirement that we recover Kubo's CPA
as $n\rightarrow 0$ and the correct atomic limit as $t_{ij}\rightarrow 0$.
Owing to the spin symmetry of $H$ we only
need to derive an equation for $G_{\u}$, and the equation for $G_{\dd}$
follows immediately.

We calculate the atomic limit Green function $g$ exactly in \sref{sAL} and
derive our many-body CPA equation for the Green function $G$ in
\sref{sCPA}. 
The CPA spectral function is studied in \sref{sSpectral}.
In the limit of infinite $J$ and $n\rightarrow 1$, Kubo \cite{rKubo2} has
introduced a CPA treatment of holes and we compare our results with this in
section \sref{shole}.  A formula for the
resistivity $\rho$ of the zero field paramagnetic state is derived in
\sref{sRho} and in \sref{sResults} $\rho$ is calculated for various
approximations to the density of states (DOS). A summary is given in
\sref{sSummary}.

\section{Atomic limit}\label{sAL}

Since we require our approximation for $G$ to be exact in the atomic limit
we must first derive $G$ in this limit. We define the retarded Green
function for operators $A$ and $B$ with no explicit time-dependence by
$\llas A\,;B\,\rras_t=$ $-\i\,\theta^{+}(t)\las\{A(t),B\}\ras$ and its
Fourier transform by $\llas A\,;B\,\rras_{\en}=
\int_{-\infty}^{\infty}{\rm d}t\,\e^{\i\en t}\llas A\,;B\,\rras_t$. Here
$\theta^+(x)=$ 1 for $x>0$ and 0 otherwise, $\en$ is restricted to the
upper half of the complex plane, and $[,]$ and $\{,\}$ are the commutator
and
anticommutator respectively. The equation of motion for the latter
Green function is
\begin{equation}\label{eMotion}
\en\lla A\,;B\,\rra_{\en}=\la\left\{A,B\right\}\ra+
\lla\left[A,H\right];B\,\rra_{\en}.
\end{equation}
The $\s$-spin one-electron Green function is given by $G^{ij}_{\s}(\en)=
\llas c_{i\s}\uu\,;c^{\dag}_{j\s}\rras_{\en}$, and we define $G_{\s}\uu=
G^{ii}_{\s}$ and $G_{\k\s}=N^{-1}\sum_{\R_{ij}}\exp(\i\k\cdot\R_{ij})
G^{ij}_{\s}$ where $\R_{ij}$ is the vector from the $j$\th to the $i$\th
lattice
site and $N$ is the number of sites. The occupation number operators are
$n_{i\s}\uu =c^{\dag}_{i\s}c_{i\s}\uu$ and we introduce the notation
$n^{\alpha}_{i\s}$, with $\alpha=\pm$, such that $n_{i\s}^{+}= n_{i\s}\uu$
and
$n_{i\s}^{-}= 1- n_{i\s}\uu$. We also define the total occupation number
$n_i=n_{i\up}+n_{i\down}$ and $n_{i}^{+}= n_{i}\uu$,\
$n_{i}^{-}= 1- n_{i}\uu$. We denote the Green function $G_{\s}$ in
the atomic limit $t_{ij}=0$ by $g_{\s}$.

In the atomic limit the eigenstates and energy eigenvalues of $H$ are
completely defined in terms of those of the single-site Hamiltonian. In
order of increasing energy these consist of $2(S+1/2)+1=2S+2$ singly
occupied states with electron spin and local spin parallel having energy 
$-JS/2$,\,\ $2S+1$ unoccupied states and $2S+1$ doubly occupied states with
energy 0, and $2(S-1/2)+1=2S$ singly occupied states with electron spin and
local spin antiparallel having energy $J(S+1)/2$. $g_{\up}$ is easily
obtained using the equation of motion method. We define $g^{\alpha}_{\up}=$
$\llas n^{\alpha}c_{\up}\uu ; c^{\dag}_{\up}\rras$,\ $s^{\alpha}_{\up}=$
$\llas n^{\alpha}S^z c_{\up}\uu ;c^{\dag}_{\up}\rras$, and
$t^{\alpha}_{\up}=\llas n^{\alpha}S^- c_{\down}\uu ;c^{\dag}_{\up}\rras$,
dropping site indices since the site referred to is always the same.
$S^{\pm}$ and $\s^{\pm}$ are the raising/lowering operators for the local
spins and conduction electrons respectively. We start with
$g^{\alpha}$ and write equations of motion for undetermined Green functions
until the system of equations closes, obtaining
\numparts
\begin{eqnarray}
\fl \en\,g^{\alpha}_{\u}(\en) = \la n_{\down}^{\alpha}\ra-
\frac{J}{2}\left[s_{\u}^{\alpha}(\en)+t^{\alpha}_{\u}(\en)\right]
\label{emotal1}\\
\fl \en\left[s^{\alpha}_{\u}(\en)+t^{\alpha}_{\u}(\en)\right] =
\la S^z n^{\alpha}_{\down}-\alpha\,{S^-}{\s^+}\ra-\frac{J\alpha}{2}
\left[s^{\alpha}_{\u}(\en)+t^{\alpha}_{\u}(\en)\right]-\frac{JS(S+1)}{2}
g^{\alpha}_{\u}(\en).
\label{emotal2}
\end{eqnarray}
\endnumparts

Now $g_{\u}=\sum_{\alpha=\pm}g^{\alpha}_{\u}$, and solving \eref{emotal1}
and \eref{emotal2} for $g^{\alpha}$ we obtain
\begin{eqnarray}
\fl g_{\u}(\en) =\frac{1}{2S+1}\left[
\frac{\las (S+S^z)n_{\down}-S^-\s^+\ras}{\en+J(S+1)/2}+
\frac{\las (S-S^z)(1-n_{\down})-S^-\s^+\ras}{\en-J(S+1)/2}\right.
\nonumber\\
\left.+\frac{\las(S+1+S^z)(1-n_{\down})+S^-\s^+\ras}{\en+JS/2}+
\frac{\las(S+1-S^z)n_{\down}+S^-\s^+\ras}{\en-JS/2}\right].
\label{edwardsal}
\end{eqnarray}
It should be noted that for finite $S$ there are four peaks in the spectral
function at energies
$\pm J(S+1)/2$ and $\pm JS/2$; for $S\rightarrow\infty$ the
upper and lower pairs of peaks merge leaving the familiar double-peaked
spectral function.
To obtain $g_{\down}(\en)$ from \eref{edwardsal} we make the changes
$n_{\s}\mapsto n_{-\s}$,\ $S^z\mapsto -S^z$,\ $S^{\pm}\mapsto S^{\mp}$,\
${\s}^{\pm}\mapsto{\s}^{\mp}$.
The weight summed over spin in the lowest band is
$(nS-2\las{\S}\cdot{\bs}\ras)/(2S+1)$, and since $\las{\S}\cdot{\bs}\ras$
$\rightarrow nS/2$ as the temperature $T\rightarrow 0$ for sufficiently large
$J$ the peak at $-J(S+1)/2$ will have 
very little weight at reasonable temperatures. Hence it will often be a good
approximation to neglect the lowest band.
The total weight summed over spin in the
lowest two bands is $(2S+2-n)/(2S+1)$, so it is clear that a $t_{ij}\ne 0$
theory that becomes exact as $t_{ij}\rightarrow 0$ will give an insulator at
half-filling.

\section{CPA Green function}\label{sCPA}

Our general strategy in this section follows that of Hubbard--- we will
retain occupation numbers as operators rather than replacing them with
their averages, but their commutators with $H_0$ will be neglected where
appropriate, corresponding to the frozen electron approximation of the
alloy analogy. Terms which are not
multiplied by $t_{ij}$ must be treated exactly in order to obtain the correct
atomic limit.

We split $G^{ij}_{\u}$ into two components, $G^{ij}_{\u}=$
$\sum_{\alpha=\pm}G^{ij\alpha}_{\u}$ where $G^{ij\alpha}_{\u}(\en)=$
$\llas n^{\alpha}_{i}c_{i\u}\uu\,;c^{\dag}_{j\u}\rras_{\en}$, corresponding
to propagation through singly ($\alpha=-$) and doubly ($\alpha=+$) occupied
sites. Neglecting $[n^{\alpha}_{i\dd},H_0]$ the equation of motion for
each 
component is then given by
\begin{eqnarray}
\fl\en\, G^{ij\alpha}_{\u}(\en)\approx
\la n^{\alpha}_{i\dd}\ra
\left(\delta_{ij}+\sum_k t_{ik}G^{kj}_{\u}(\en)\right)
+\sum_k t_{ik}\lla \delta n^{\alpha}_{i\dd}
c\uu_{k\u}\,;c_{j\u}^{\dag}\rra_{\en}
\nonumber\\
-\frac{J}{2}\left(S^{ij\alpha}_{\up}(\en)+T^{ij\alpha}_{\up}(\en)\right)
\label{ecpa1}
\end{eqnarray}
where we have introduced the Green functions $S^{ij\alpha}_{\up}(\en)=$
$\llas n^{\alpha}_{i}S^{z}_i c_{i\up}\uu\,;c^{\dag}_{j\up}\rras_{\en}$ and
$T^{ij\alpha}_{\up}(\en)=$ $\llas n^{\alpha}_{i}S^{-}_{i}$
$c_{i\down}\uu\,;$ $c^{\dag}_{j\up}\rras_{\en}$
and the notation $\delta A=A-\las A\ras$ for any operator $A$.
The second term on the
right-hand side of \eref{ecpa1} is Hubbard's scattering correction in which
the deviation of $n^{\alpha}_{i\dd}$ from its average is accounted for. The
last term of \eref{ecpa1} is more complicated than in the case of the
Hubbard model, containing the as yet undetermined Green functions $S_{\u}$ 
and $T_{\u}$. These correspond respectively to propagation of the electron 
as an $\u$-spin and, following spin-flip scattering from a local spin,
as a $\dd$-spin; the presence of $T_{\u}$ will couple the equations for
$G_{\u}$ and $G_{\dd}$.

We first treat the scattering correction, splitting the relevant Green
function into two components, $\llas \delta n^{\alpha}_{i\dd}c\uu_{k\u}\,;
c_{j\u}^{\dag}\rras_{\en}=\sum_{\beta=\pm}\llas\delta n^{\alpha}_{i\dd}
n^{\beta}_{k}c\uu_{k\u}\,;c_{j\u}^{\dag}\rras_{\en}$. It is assumed that
$t_{ii}=0$, so from \eref{ecpa1} it may be seen that this Green function is
needed only for $i\ne k$. The equations of motion are, for $i\ne k$,
\begin{eqnarray}
\fl\en\lla\delta n^{\alpha}_{i\dd}n^{\beta}_{k} c\uu_{k\u}\,;
c_{j\u}^{\dag}\rra_{\en} \approx
\la\delta n^{\alpha}_{i\dd}\delta n^{\beta}_{k\dd}\ra\delta_{jk}
+\sum_l t_{kl} \lla\delta n^{\alpha}_{i\dd}n^{\beta}_{k\dd}c\uu_{l\u}\,;
c_{j\u}^{\dag}\rra_{\en}
\nonumber\\
 -\frac{J}{2} \lla\delta n^{\alpha}_{i\dd}n^{\beta}_{k}
\left(S^{z}_k c\uu_{k\u}+S^{-}_{k}c\uu_{k\dd}\right);
c^{\dag}_{j\u}\rra_{\en}
\label{esc1}
\end{eqnarray}
where $[n^{\beta}_{k\dd},H_0]$ and all commutators involving
$\delta n^{\alpha}_{i\dd}$ have been neglected. This is consistent with the
strategy stated above. As a further approximation the first term on the
right-hand side of \eref{esc1}, a two-site correlation function, is
dropped and we set $n^{\beta}_{k\dd}\approx\las n^{\beta}_{k\dd}\ras$ in
the second term, which corresponds to neglecting a second scattering
correction.

The system of equations for the scattering correction is now closed apart
from the last Green function in \eref{esc1}. In the equation of motion for 
this term we use the fact that
\begin{equation}
n^{\beta}_{k}(S^{z}_k c\uu_{k\u}+S^{-}_{k}c\uu_{k\dd})=
\cases{
(S^{z}_{k}n_{k\dd}^{\beta}-\beta S^{-}_{k}\s^{+}_{k})c\uu_{k\u}
&for $\beta=+$\\
c\uu_{k\u}(S^{z}_{k}n_{k\dd}^{\beta}-\beta S^{-}_{k}\s^{+}_{k})
&for $\beta=-$\\}
\end{equation}
and replace $S^{z}_{k}n_{k\dd}^{\beta}-\beta S^{-}_{k}\s^{+}_{k}$ by its
average in the Green function coming from $H_0$. Neglecting two-site
correlation functions and commutators of  $\delta n^{\alpha}_{i\dd}$ we
obtain the equation of motion, for $i\ne k$,
\begin{eqnarray}
\fl\en\lla\delta n^{\alpha}_{i\dd}n^{\beta}_{k}
\left(S^{z}_k c_{k\u}\uu+S^{-}_{k}c_{k\dd}\uu\right);
c^{\dag}_{j\u}\rra_{\en}\approx
\la S^{z}_{k}n^{\beta}_{k\dd}-\beta S^{-}_{k}\s^{+}_{k}\ra
\sum_{l}t_{kl}\lla \delta n^{\alpha}_{i\dd}c_{l\u}\uu\,;
c^{\dag}_{j\u}\rra_{\en}
\nonumber\\
-\frac{J\beta}{2}\lla\delta n^{\alpha}_{i\dd}
n^{\beta}_{k}\left(S^{z}_{k}
c_{k\u}\uu+S^{-}_{k}c_{k\dd}\uu\right);c^{\dag}_{j\u}\rra_{\en}
\nonumber\\
-\frac{JS(S+1)}{2}\lla\delta n^{\alpha}_{i\dd}
n^{\beta}_{k}c_{k\u}\uu;c^{\dag}_{j\u}\rra_{\en},
\label{esc2}
\end{eqnarray}
thus closing the system of equations for the scattering correction.

We solve \eref{esc2} for the Green function on the left-hand side,
substitute the result into \eref{esc1}, and rearrange and sum over $\beta$
to obtain, again for $i\ne k$,
\begin{equation}
\lla\delta n^{\alpha}_{i\dd}c_{k\up}\uu\,;c_{j\up}^{\dag}\rra_{\en}=
g_{\up}(\en)\sum_l t_{kl}
\lla\delta n^{\alpha}_{i\dd}c_{l\up}\uu\,;c_{j\up}^{\dag}\rra_{\en},
\label{eHubbard}
\end{equation}
where $g_{\u}(\en)$ is the atomic-limit Green function presented in
\eref{edwardsal}. In the appendix of \cite{rHubbard3} Hubbard solved this
equation in terms of $\llas\delta n^{\alpha}_{i\dd}c_{i\up}\uu\,;
c_{j\up}^{\dag}\,\rras_{\en}$:
\begin{equation}
\lla\delta n^{\alpha}_{i\dd}c_{k\up}\uu\,;c_{j\up}^{\dag}\rra_{\en}=
\left(\sum_{l}W_{kl}^{i\up}(\en)t_{li}\right)
\lla\delta n^{\alpha}_{i\dd}c_{i\up}\uu\,;c_{j\up}^{\dag}\rra_{\en},
\quad (i\ne k)\label{eHubbardSol}
\end{equation}
where $W_{kl}^{i\s}$ is defined by
\numparts
\begin{eqnarray}
W_{kl}^{i\s}(\en)={\tilde g}^{kl}_{\s}(\en)-
\frac{{\tilde g}^{ki}_{\s}(\en){\tilde g}^{il}_{\s}(\en)}
{{\tilde g}^{ii}_{\s}(\en)}\\ \label{GandT}
{\tilde g}^{ij}_{\s}(\en)=\frac{1}{N}\sum_{\k}
\frac{\exp(\i\k\cdot\R_{ij})}{g_{\s}(\en)^{-1}-t_{\k}}.
\end{eqnarray}
\endnumparts
It is easy to check Hubbard's solution: from \eref{GandT} it may
be shown that $g_{\s}\sum_k t_{ik}{\tilde g}^{kj}_{\s}={\tilde g}^{ij}_{\s}
-g_{\s}\delta_{ij}$ and hence that $g_{\s}\sum_l t_{kl}W_{lj}^{i\s}=
W_{kj}^{i\s}-g_{\s}\delta_{kj}$ for $i\ne k$, and this can be used to
verify that substituting \eref{eHubbardSol} into \eref{eHubbard} does
indeed give a solution.

As in Hubbard's case ${\tilde g}^{ij}_{\s}$ is a zeroth order approximation
to $G^{ij}_{\s}$ in which both scattering and resonance broadening
corrections are neglected, and following Hubbard we make the
self-consistent replacement ${\tilde g}^{ij}_{\up}\mapsto G^{ij}_{\up}$;
this is the essential self-consistency step of the CPA. 
From \eref{eHubbardSol} the scattering correction term in
\eref{ecpa1} is given by
\begin{equation}
\sum_k t_{ik} \lla\delta n^{\alpha}_{i\dd}c_{k\up}\uu\,;
c_{j\up}^{\dag}\rra_{\en}=J_{\up}(\en)
\lla\delta n^{\alpha}_{i\dd}c_{i\up}\uu\,;c_{j\up}^{\dag}\rra_{\en}
\label{escatt}
\end{equation}
where $J_{\s}(\en)=\sum_{kl}t_{ik}W_{kl}^{i\s}(\en)t_{li}$. A result that
will be useful later is
\begin{equation}
J_{\s}(\en)=\en-\Sigma_{\s}(\en)-G_{\s}(\en)^{-1}
\label{eJ}
\end{equation}
which holds
if the self-energy $\Sigma_{\s}(\en)$ is local--- as is expected for a CPA.
This result may easily be established by making Fourier transforms and using
$G^{ij}_{\s}(\en)=N^{-1}\sum_{\k}\exp(i\k\cdot\R_{ij})G_{\k\s}(\en)$. We now
also define the useful quantity
\begin{equation}
E_{\s}(\en)=\en-J_{\s}(\en).
\label{enew}
\end{equation}

It remains to find expressions for the unknown Green functions $S_{\u}$ and
$T_{\u}$ in \eref{ecpa1}; since these represent propagation as $\u$- and
$\dd$-spins respectively we neglect the commutators $[n^{\alpha}_{i\dd},
H_0]$ and $[n^{\alpha}_{i\u},H_0]$ in the respective equations of motion:
\begin{eqnarray} 
\fl\en\,S^{ij\alpha}_{\u}(\en)\approx
\la S^{z}_{i}n^{\alpha}_{i\down}\ra\left(\delta_{ij}+
\sum_{k}t_{ik}G^{kj}_{\u}(\en)\right)
+\sum_{k} t_{ik} \lla \delta(n^{\alpha}_{i\down}S^{z}_i)c_{k\up}\uu\,;
c^{\dag}_{j\up}\rra_{\en}
-\frac{J}{2}\delta_{\alpha +}T^{ij\alpha}_{\u}(\en)
\nonumber\\
-\frac{J}{2}\lla n^{\alpha}_{i}\left(({S^{z}_{i}})^2 c_{i\up}\uu
+S^{z}_{i}S^{-}_{i}c_{i\down}\uu\right);c^{\dag}_{j\up}\rra_{\en}
\label{esp}
\end{eqnarray}
\begin{eqnarray}
\fl\en\, T^{ij\alpha}_{\u}(\en)\approx
-\alpha\la S^{-}_{i}\s^{+}_{i}\ra\left(\delta_{ij}+
\sum_{k}t_{ik}G^{kj}_{\u}(\en)\right)
\nonumber\\
+\sum_{k}t_{ik}\lla \left(n^{\alpha}_{i\up}S^{-}_{i}c_{k\down}\uu+
\alpha\la S^{-}_{i}\s^{+}_{i}\ra c_{k\up}\uu\right);c^{\dag}_{j\up}
\rra_{\en}
\nonumber\\
-\frac{J}{2}\left(\alpha S^{ij\alpha}_{\u}(\en)-\delta_{\alpha -}
T^{ij\alpha}_{\u}(\en)+S(S+1)G^{ij\alpha}_{\u}(\en)\right)
\nonumber\\
+\frac{J}{2}\lla n^{\alpha}_{i}\left(({S^{z}_{i}})^2 c_{i\up}\uu
+S^{z}_{i}S^{-}_{i}c_{i\down}\uu\right);c^{\dag}_{j\up}\rra_{\en}.
\label{esm}
\end{eqnarray}
The second terms on the right-hand sides of these equations are the
scattering corrections; the scattering correction in \eref{esp} is of the same
form as in \eref{ecpa1}, but in \eref{esm} it is
more complicated. It is not clear what the average `zeroth order'
Green function should be here, but we have chosen it to be
$-\alpha\las S^{-}_{i}\s^{+}_{i}\ras G^{kj}_{\u}(\en)$ as this
makes the first terms on the right-hand sides
of \eref{ecpa1}, \eref{esp}, and \eref{esm} all of
the same form; it turns out that this is necessary for the consistency of
the approximation. The last Green function in \eref{esp} and \eref{esm}
involves higher order spin operators and is in general unknown;
this term must be treated using an approximation which is exact in
the atomic limit.

The scattering correction of \eref{esp} may be treated in the same way as
that of \eref{ecpa1}; the only difference is that
$\delta n_{i\down}^{\alpha}$ is replaced by $\delta(n^{\alpha}_{i\down}
S^z_i)$, hence
\begin{equation}
\sum_{k} t_{ik} \lla \delta (n^{\alpha}_{i\down}S^{z}_i)
c_{k\up}\uu\,;c^{\dag}_{j\up}\rra_{\en}
\approx
J_{\up}(\en)\lla\delta(n^{\alpha}_{i\down}S^{z}_{i})c_{i\up}\uu\,;
c^{\dag}_{j\up}\rra_{\en}.\label{esc1a}
\end{equation}
It is less obvious how to treat the scattering correction of \eref{esm}.
Instead of giving a rigorous derivation we proceed by analogy and make the 
apparently reasonable approximation
\numparts
\begin{eqnarray}
\fl\sum_{k}t_{ik}\lla \left(n^{\alpha}_{i\up}S^{-}_{i}c_{k\down}\uu+
\alpha\la S^{-}_{i}\s^{+}_{i}\ra c_{k\up}\uu\right);c^{\dag}_{j\up}
\rra_{\en}
\nonumber\\
\lo{\approx}
J_{\dd}(\en)\lla n^{\alpha}_{i\up}S^{-}_{i}c_{i\down}\uu;c^{\dag}_{j\up}
\rra_{\en}+ J_{\up}(\en)\lla\alpha\la S^{-}_{i}\s^{+}_{i}\ra c_{i\up}\uu
;c^{\dag}_{j\up}\rra_{\en}
\\
\lo= J_{\dd}(\en)T^{ij\alpha}_{\u}(\en)
+\alpha\la S^{-}_{i}\s^{+}_{i}\ra J_{\up}(\en)G^{ij}_{\u}(\en).
\label{esc2a}
\end{eqnarray}
\endnumparts
It will be seen later that this is consistent with the rest of the
derivation, leading for instance to a local self-energy as
expected in a CPA.

The system of equations of motion is now closed apart from the Green
function $\llas n^{\alpha}_{i}(({S^{z}_{i}})^2 c_{i\up}\uu+S^{z}_{i}
S^{-}_{i}c_{i\down}\uu);c^{\dag}_{j\up}\rras_{\en}$ appearing
in \eref{esp} and \eref{esm}. In general we will have to write more
equations of motion to close the system. This is
straightforward
to do if we make approximations analogous to those used to obtain the 
equations of motion for $S_{\u}$ and $T_{\u}$. Then the Green functions
$\llas n^{\alpha}_{i}(S^{z}_i)^m c_{i\up}\uu\,;c^{\dag}_{j\up}\rras_{\en}$ 
and $\llas n^{\alpha}_{i}(S^z_i)^{m-1}S^{-}_{i}c_{i\down}\uu\,;
c^{\dag}_{j\up}\rras_{\en}$ and the expectations
$\las (S^z_i)^m n^-_{i\dd}\ras$ and $\las (S^z_i)^{m-1}S^-_i\s^+_i\ras$
for $m=1,
\ldots,\ 2S$ are brought into the system of equations. The resulting system
of $4S+1$ equations per spin is complicated to solve self-consistently for 
large $S$ however, so we will restrict ourselves to simple special cases in
which we do not need any more equations of motion; in the next three
subsections we will consider the paramagnetic state in zero magnetic field 
for arbitrary $S$, the case $S=1/2$ for arbitrary magnetization, and the
case of saturated ferromagnetism for arbitrary $S$.

Our approximation scheme is now complete and is entirely self-consistent,
i.e.\ all the expectations appearing in the system of equations can be
obtained from the Green functions in the system via the
relation \cite{rRKubo}
\begin{equation}
\la BA\,\ra=-\int_{-\infty}^{\infty}\frac{\d\en}{\pi}
f(\en-\mu)\,{\rm Im}\lla A\,;B\,\rra_{\en}
\label{esc}
\end{equation}
where $f(\en)=1/(1+\exp(\beta\en))$ is the Fermi function and $\mu$ is the
chemical potential. Also, since all the approximations made above have
been in terms proportional to $t_{ij}$
the approximation is exact in the atomic 
limit for all $S$ and $n$ as required.
The system is assumed to be homogeneous, so we now drop the
site indices of the expectations.

\subsection{Paramagnetism}\label{ssPara}

In the case of paramagnetism with zero field $J_{\u}=J_{\dd}$. Hence
after substituting \eref{esc1a} and \eref{esc2a} in \eref{esp} and \eref{esm},
respectively, we can add the resulting equations so that $S^{ij\alpha}$ and
$T^{ij\alpha}$ occur only in the combination
$S^{ij\alpha}+T^{ij\alpha}$:
\begin{eqnarray}
\fl\en\, \left(S^{ij\alpha}(\en)+T^{ij\alpha}(\en)\right) =
\la S^{z}n^{\alpha}_{\down}-\alpha S^{-}\s^{+}\ra
\left(\delta_{ij}+\sum_{k}t_{ik}G^{kj}(\en)
-J(\en)\,G^{ij}(\en)\right)
\nonumber\\ 
+\left(J(\en)-\frac{J\alpha}{2}\right)
\left(S^{ij\alpha}(\en)+T^{ij\alpha}(\en)\right)
-\frac{JS(S+1)}{2}G^{ij\alpha}(\en).
\label{epara}
\end{eqnarray}
Since $S^{ij\alpha}$ and $T^{ij\alpha}$ also enter \eref{ecpa1} as 
$S^{ij\alpha}+T^{ij\alpha}$ we have now closed the system.
We substitute \eref{escatt} and \eref{epara} into \eref{ecpa1}
and rearrange to obtain
\begin{equation}
G^{ij}(\en)=\left(\delta_{ij}+\sum_k t_{ik}G^{kj}(\en)- J(\en) G^{ij}(\en)
\right){\tilde G}(\en)
\label{paraGF}
\end{equation}
where ${\tilde G}$ is defined by
\begin{equation}
{\tilde G}(\en)=\sum_{\alpha=\pm}\frac{(E(\en)+\alpha J/2)(n/2)^{\alpha}
+J\alpha/2\,\las \S\cdot\bs\ras}
{\left(E(\en)-\alpha JS/2\right)\left(E(\en)+\alpha J(S+1)/2\right)},
\label{eparaGF}
\end{equation}
$E(\en)=\en-J(\en)$, and $(n/2)^{\alpha}=\delta_{\alpha-}+\alpha n/2$.
The spin symmetry of the paramagnetic state has been used
to simplify the expectations appearing in \eref{eparaGF}.

Taking Fourier transforms and solving for $G_{\k}$
\eref{paraGF} becomes
\begin{equation}
G_{\k}(\en)=\frac{1}{J(\en)+{\tilde G}(\en)^{-1}-t_{\k}}.
\label{eGkp}
\end{equation}
Since in general $G_{\k}(\en)=(\en-t_{\k}-\Sigma_{\k}(\en))^{-1}$ this
implies that the self-energy is local, $\Sigma_{\k}(\en)=\Sigma(\en)$,
and so using expression \eref{eJ} for $J(\en)$ \eref{eGkp} becomes
\begin{equation}
G_{\k}(\en)=\frac{1}{\en-t_{\k}-\Sigma(\en)+{\tilde G}(\en)^{-1}-
G(\en)^{-1}},
\label{eGkp1}
\end{equation}
and for consistency in this equation we must have $G={\tilde G}$.
The existence of this simple self-consistency condition
defining $G$ is our main justification for the approximations made for the
scattering correction of \eref{esm}. If the approximations are changed we
will not obtain a consistent expression defining a local self-energy.

In the low density limit $n\rightarrow 0$ \eref{eparaGF} reduces to
\begin{equation}
G(\en)= \frac{E(\en)-J/2}{(E(\en)+JS/2)(E(\en)-J(S+1)/2)},
\end{equation}
which is Kubo's \cite{rKubo} equation for the paramagnetic state Green
function as required.

\subsection{The $S=1/2$ case for arbitrary magnetization}\label{ssHalf}

For $S=1/2$ we have $(S^{z}_{i})^2=1/4$ and
$S^{z}_{i}S^{-}_{i}=-1/2\,S^{-}_{i}$ for arbitrary magnetization, so the
last Green function in \eref{esp} and \eref{esm} may be simplified:
\begin{equation}
\lla n^{\alpha}_{i}\left(({S^{z}_{i}})^2 c_{i\up}\uu
+S^{z}_{i}S^{-}_{i}c_{i\down}\uu\right);c^{\dag}_{j\up}\rra_{\en}=
\frac{1}{4}G^{ij\alpha}_{\u}(\en)-\frac{1}{2}T^{ij\alpha}_{\u}(\en),
\end{equation}
closing the system.
Substituting the scattering corrections \eref{escatt}, \eref{esc1a}, and
\eref{esc2a} into the equations of motion
\eref{ecpa1}, \eref{esp}, and \eref{esm} the system of equations reduces to
\numparts
\begin{eqnarray}
\label{egf1}
E_{\up}\uu G^{ij\alpha}_{\up} = 
\las n^{\alpha}_{\down}\ras\lambda^{ij}_{\u}
-\frac{J}{2}\left(S^{ij\alpha}_{\up}
+T^{ij\alpha}_{\up}\right)\\
E_{\up}\uu S^{ij\alpha}_{\up} = \las S^z n^{\alpha}_{\down}\ras
\lambda^{ij}_{\u}-\frac{J}{8}
G^{ij\alpha}_{\up}-\frac{J\alpha}{4}T^{ij\alpha}_{\up}
\label{egf2}\\
E^{\alpha}_{\down}T^{ij\alpha}_{\up} = -\alpha\las S^- \s^+\ras
\lambda^{ij}_{\u}
-\frac{J}{4}
G^{ij\alpha}_{\up}- \frac{J\alpha}{2} S^{ij\alpha}_{\up},\label{egf3}
\end{eqnarray}
\endnumparts
where
\begin{equation}
\lambda^{ij}_{\s}(\en)= \delta_{ij}+\sum_k t_{ik}G^{kj}_{\s}(\en)-
J_{\s}(\en)G^{ij}_{\s}(\en),
\end{equation}
$E_{\s}(\en)=\en-J_{\s}(\en)$, and $E^{\alpha}_{\s}=E_{\s}+J\alpha/4$.
These equations can be solved using a similar method to that of the
previous subsection, with an analogous self-consistency condition
occurring, and the local Green function is given by 
\begin{equation}
\fl G_{\up}(\en)=
\sum_{\alpha=\pm}\frac{\las n^{\alpha}_{\down}\ras
\left(E_{\up}(\en) E_{\down}^{\alpha}(\en)-J^2/8\right)
-J/2\,\left(\las S^{z} n^{\alpha}_{\down}\ras E^{-\alpha}_{\down}(\en)
-\alpha\,\las S^{-}{\s}^{+}\ras E^{-\alpha}_{\up}(\en)\right)}
{E^{-\alpha}_{\up}(\en)\left(
E_{\up}^{\alpha}(\en) E_{\down}^{\alpha}(\en)-J^2/4\right)}.
\label{egfs}
\end{equation}
The expectations $\las n^{\alpha}_{\down}\ras$,
$\las S^{z} n^{\alpha}_{\down}\ras$, and $\las S^{-}{\s}^{+}\ras$ may be
calculated self-consistently using \eref{egf1}--\eref{egf3} and \eref{esc}.

If we set $n=0$ \eref{egfs} reduces to
\begin{equation}
G_{\up}(\en)=\frac{E_{\up}(\en)E^-_{\down}(\en) -J^2/8-J/2\las S^z\ras
E_{\down}^+ (\en)}{E_{\up}^+(\en)[E_{\up}^-(\en)E_{\down}^-(\en)-J^2/4]}.
\label{eShk}
\end{equation}
For $S=1/2$ the probabilities of a local spin being up or down are given by
$P(S^z=\pm 1/2)=1/2\pm\las S^z\ras$, and this fact may be used to show that
\eref{eShk} is equal to Kubo's equation \cite{rKubo}
for $G_{\u}$ in the $S=1/2$ case. Agreement with Kubo's approximation in
the
case of arbitrary magnetization is a much more stringent condition on our
CPA than agreement just in the paramagnetic case, so this gives us
confidence
that our approximation is indeed a many-body extension of Kubo's.

\subsection{Saturated ferromagnetism for arbitrary $S$}\label{ssFerro}

If we assume the existence of a
saturated ferromagnetic state where all local spins and conduction
electron spins are aligned parallel to the positive $z$-axis
substantial simplification occurs. The
expectations occurring in the $\u$- and $\dd$-spin systems are given by
\numparts
\begin{eqnarray}
\fl
\la n^{\alpha}_{\down}\ra=\delta_{\alpha-}\qquad&
\la S^z n^{\alpha}_{\down}\ra=S\delta_{\alpha-}\qquad&
\la S^- \s^+\ra=0\\
\fl
\la n^{\alpha}_{\up}\ra=\delta_{\alpha-}+\alpha n\qquad&
\la -S^z n^{\alpha}_{\up}\ra=-S\left(\delta_{\alpha-}+\alpha n\right)
\qquad&
\la S^+ \s^-\ra=0,
\end{eqnarray}
\endnumparts
and the Green functions $G_{\s}$, $S_{\s}$, and $T_{\s}$
may be simplified using the relations
\numparts
\begin{eqnarray}
G^{ij+}_{\up}=T^{ij\alpha}_{\up}=0 \qquad&
S^{ij\alpha}_{\u}=S\delta_{\alpha-} G^{ij}_{\u}\\
T^{ij+}_{\dd}=0 \qquad& S^{ij\alpha}_{\dd}=-S G^{ij\alpha}_{\dd}.
\end{eqnarray}
\endnumparts
The higher order Green functions are given by
\numparts
\begin{eqnarray}
\lla n^{\alpha}_{i}\left(({S^{z}_{i}})^2 c_{i\u}\uu
+S^{z}_{i}S^{-}_{i}c_{i\dd}\uu\right);c^{\dag}_{j\u}\rra_{\en}
= S^2\delta_{\alpha-} G^{ij}_{\u}\\
\lla n^{\alpha}_{i}\left(({S^{z}_{i}})^2 c_{i\dd}\uu
-S^{z}_{i}S^{+}_{i}c_{i\u}\uu\right);c^{\dag}_{j\dd}\rra_{\en}
=S^2 G^{ij\alpha}_{\dd} -S\delta_{\alpha-}T^{ij}_{\dd}.
\end{eqnarray}
\endnumparts
The system therefore closes without needing any more equations. After
simplification the set of equations of motion (for both spin types) reduces
 to
\numparts
\begin{eqnarray}
\left(E_{\u}+\frac{JS}{2}\right)G^{ij}_{\u}=\lambda^{ij}_{\u}\\
\left(E_{\dd}-\frac{JS}{2}\right)G^{ij+}_{\dd}=n\lambda^{ij}_{\dd}\\
\left(E_{\dd}-\frac{JS}{2}\right)G^{ij-}_{\dd}=(1-n)\lambda^{ij}_{\dd}
-\frac{J}{2}T^{ij}_{\dd}\\
\left(E_{\u}+\frac{J(S-1)}{2}\right)T^{ij}_{\dd}=-JS\, G^{ij-}_{\dd},
\end{eqnarray}
\endnumparts
with the other equations satisfied automatically.
Here $\lambda^{ij}_{\s}$ and $E_{\s}$ are defined as in the
previous subsection.

We can solve these equations in the same way as in the previous two
subsections, with the usual self-consistency condition similar to
\eref{eGkp1} applying,
and the local Green functions are given by
\numparts
\begin{eqnarray}\label{e34a}
G_{\u}=\frac{1}{E_{\u}+JS/2}\\
G_{\dd}=\frac{n}{E_{\dd}-JS/2}+\frac{1-n}{E_{\dd}-JS/2-J^2 S/2\,
(E_{\u}+J(S-1)/2)^{-1}}.\label{e34b}
\end{eqnarray}
\endnumparts
The $\u$-spin Green function here is just the free Green function shifted
in energy, as might be expected. In the $n\rightarrow 0$ limit these
equations
reduce to those of Kubo \cite{rKubo},  providing a further check on our
approximation. It is shown in the next section that these Green functions are
in fact not consistent with the initial assumption of a saturated ferromagnetic
state.

\section{CPA spectral function}\label{sSpectral}

\begin{figure}
\begin{center}
\leavevmode
\hbox{%
\epsfxsize=0.5\textwidth
\epsffile{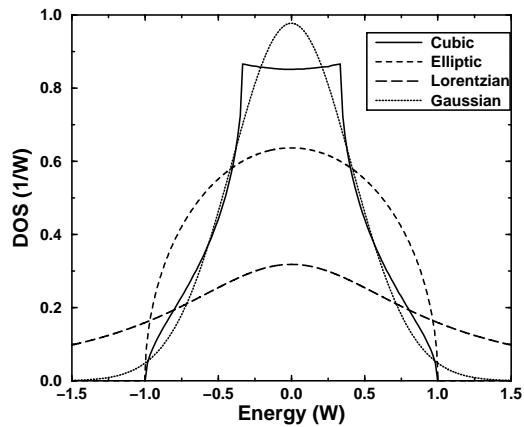}}
\end{center}
\caption{The DOS's $D_{\rm c}$, $D_{\rm e}$, $D_{\rm l}$, and
$D_{\rm g}$ plotted in units of $W$.
\label{fDOSes}}
\end{figure}

In this section we study the DOS, first for the saturated ferromagnetic
state
and then for the paramagnetic state. We make the approximation of replacing
the
true DOS $D_{\rm c}$ for a simple cubic lattice with the elliptic DOS,
$D_{\e}(\en)=2/(\pi W^2)\sqrt{W^2-\en^2}$ where $2W$ is the bandwidth.
For an elliptic DOS the free Green function is given by
$G_0(\en)= 2/W^2(\en-\sqrt{\en^2-W^2})$. Since the self-energy
$\Sigma_{\s}(\en)$ is local the full Green function is given by
$G_{\s}(\en)=G_0(\en-\Sigma_{\s}(\en))$, and it follows immediately that
$J_{\s}=W^2/4\,G_{\s}$. The Green function equations therefore become
algebraic in this approximation, greatly simplifying the
calculations. The elliptic DOS is also a fairly good approximation to the
true 
DOS, unlike the Lorentzian DOS $D_{\rm l}(\en)=(W/\pi)/(W^2+\en^2)$
considered by Furukawa \cite{rFurukawa}. In the infinite dimensional
limit of a hypercubic lattice with nearest neighbour hopping--- the
scenario
considered by Furukawa--- it may be shown that the true DOS is 
a Gaussian if the hopping $t_{ij}$
is scaled as the inverse square root of the
number of
dimensions \cite{rMuller}. With the scaling appropriate to three dimensions
 the
Gaussian DOS is given by $D_{\rm g}(\en)=(3/\pi)^{1/2}\exp[-3(\en/W)^2]/W$.
These DOS's are compared in \fref{fDOSes}.

\subsection{Saturated ferromagnetism}

\begin{figure}
\begin{center}
\leavevmode
\hbox{%
\epsfxsize=0.5\textwidth
\epsffile{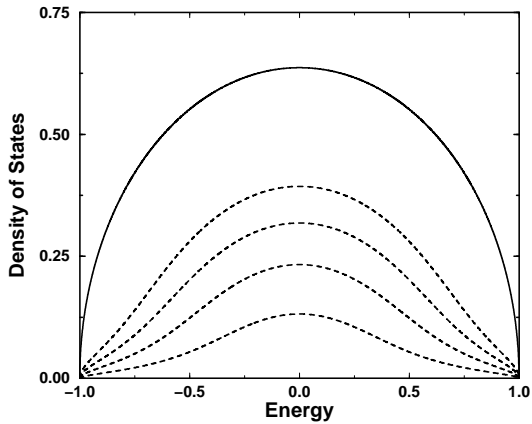}}
\end{center}
\caption{The $\u$-spin (solid curve, independent of $n$) and $\dd$-spin
(dashed curves, $n=0,\ 0.25,\ 0.5,\ 0.75$, DOS decreasing with
increasing $n$) DOS's in the saturated ferromagnetic state for $S=1/2$ and
$J=\infty$. Energy units of $W$ are used.
\label{ffdos}}
\end{figure}

We study the ferromagnetic state with complete spin alignment
in the strong-coupling $J\rightarrow\infty$
limit which is most favourable to ferromagnetism, shifting the energy
origin
by $-JS/2$ in order to have the zero of energy in the lowest band.
Equations \eref{e34a} and \eref{e34b} become
\numparts
\begin{eqnarray}
G_{\u}= 1/E_{\u}\\
G_{\dd}= \frac{1-n}{E_{\dd}+2S E_{\u}}.
\end{eqnarray}
\endnumparts
The DOS of this state is plotted in \fref{ffdos}
for $S=1/2$ and various $n$. Since the $\up$- and $\dd$-spin DOS's are
nonzero for the same range of energies it is clear that consistent saturated
ferromagnetism does not occur within our approximation for any $n$.
It may easily be shown that this is true
for any finite $S$.
 Since strong ferromagnetism is expected to occur in the DE
model for at least some parameter values this is a limitation of our
approximation and suggests that our CPA may not be very good at low
temperatures. Of course in common with the usual alloy CPA we cannot obtain
a true Fermi liquid groundstate in our CPA--- the imaginary part of the
self-energy does not vanish at the Fermi surface at $T=0$--- so we only
expect
our CPA to describe the DE model well at finite (but not necessarily large)
temperatures where this is not a problem. The possibility of weak
ferromagnetism occurring in our approximation will be discussed elsewhere;
preliminary work on the magnetic susceptibility has already
appeared in
\cite{rUs}--\cite{rMe}.

Note that the weights of the $\u$- and $\dd$-spin bands here are 1
and $(1-n)/(2S+1)$ respectively, so in the classical spin limit
$S\rightarrow\infty$ there is no possibility of $\dd$-spin weight occurring
near the Fermi level. In fact for $S\rightarrow\infty$ the saturated state is
always stable against spin reversal but for finite $S$ this is not always the
case \cite{rBruntonEdwards}.

\subsection{Paramagnetism}

\begin{figure}
\begin{center}
\leavevmode
\hbox{%
\epsfxsize=0.5\textwidth\epsffile{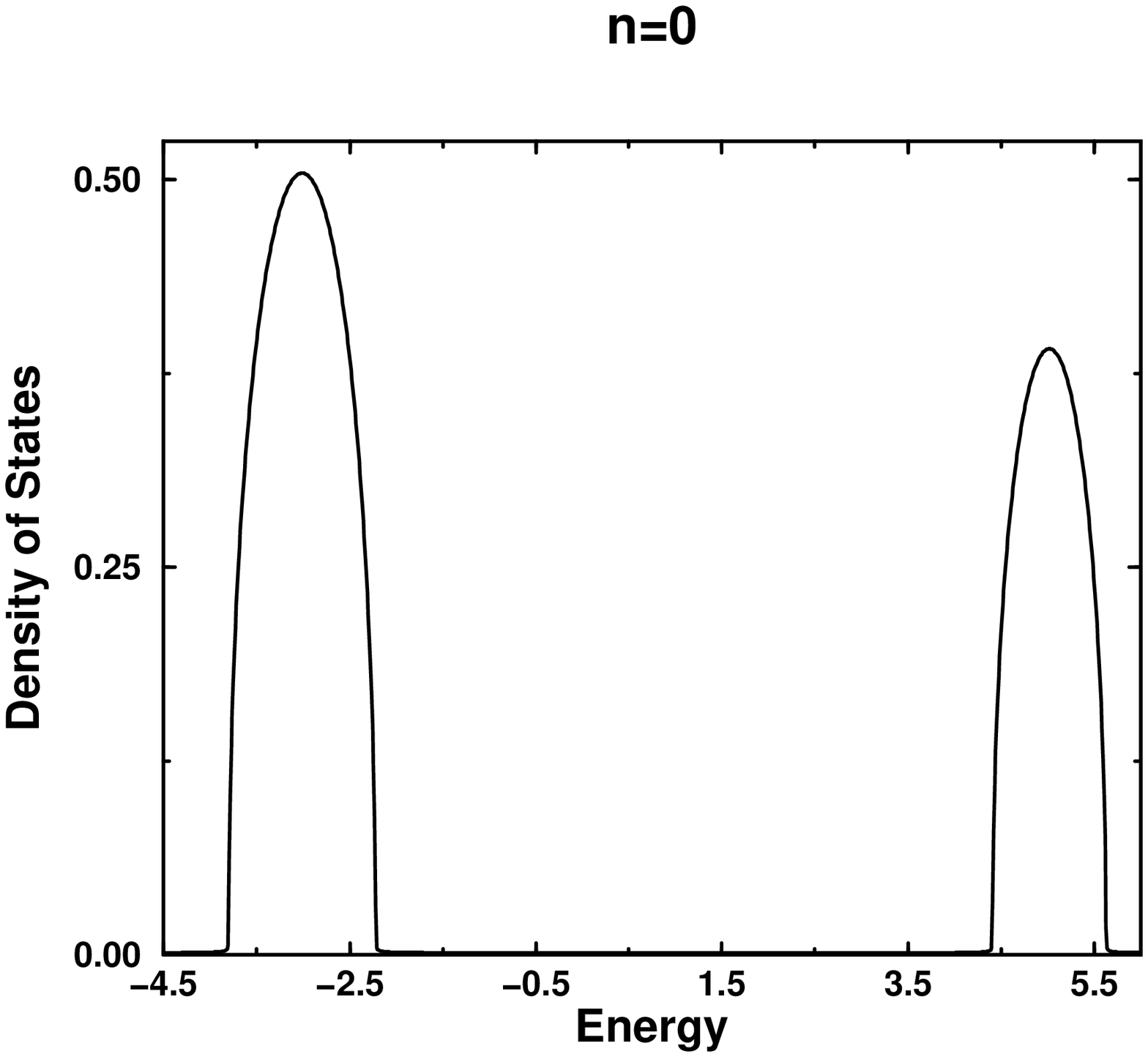}
\epsfxsize=0.5\textwidth\epsffile{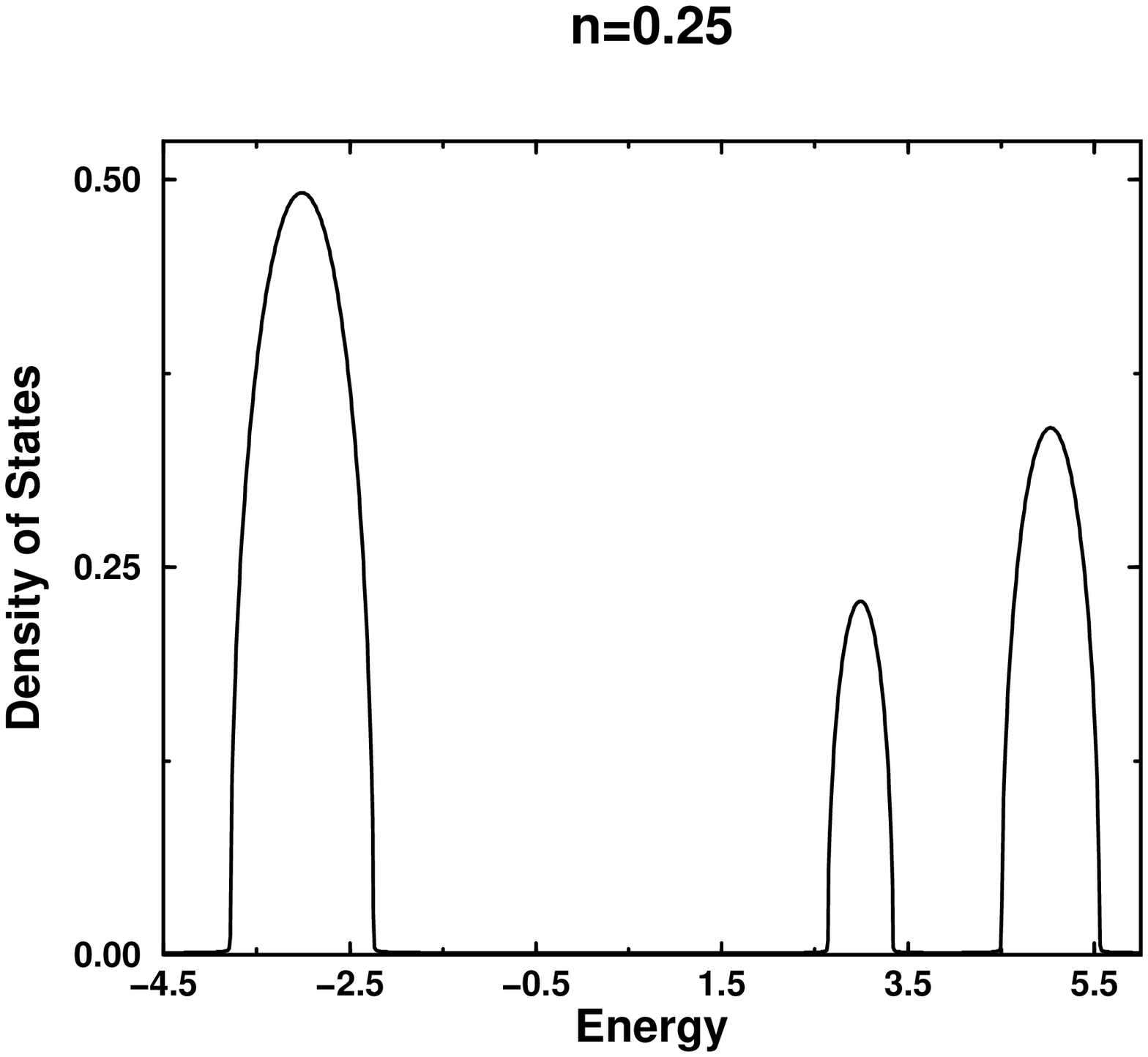}}
\vskip 0.25in
\hbox{%
\epsfxsize=0.5\textwidth\epsffile{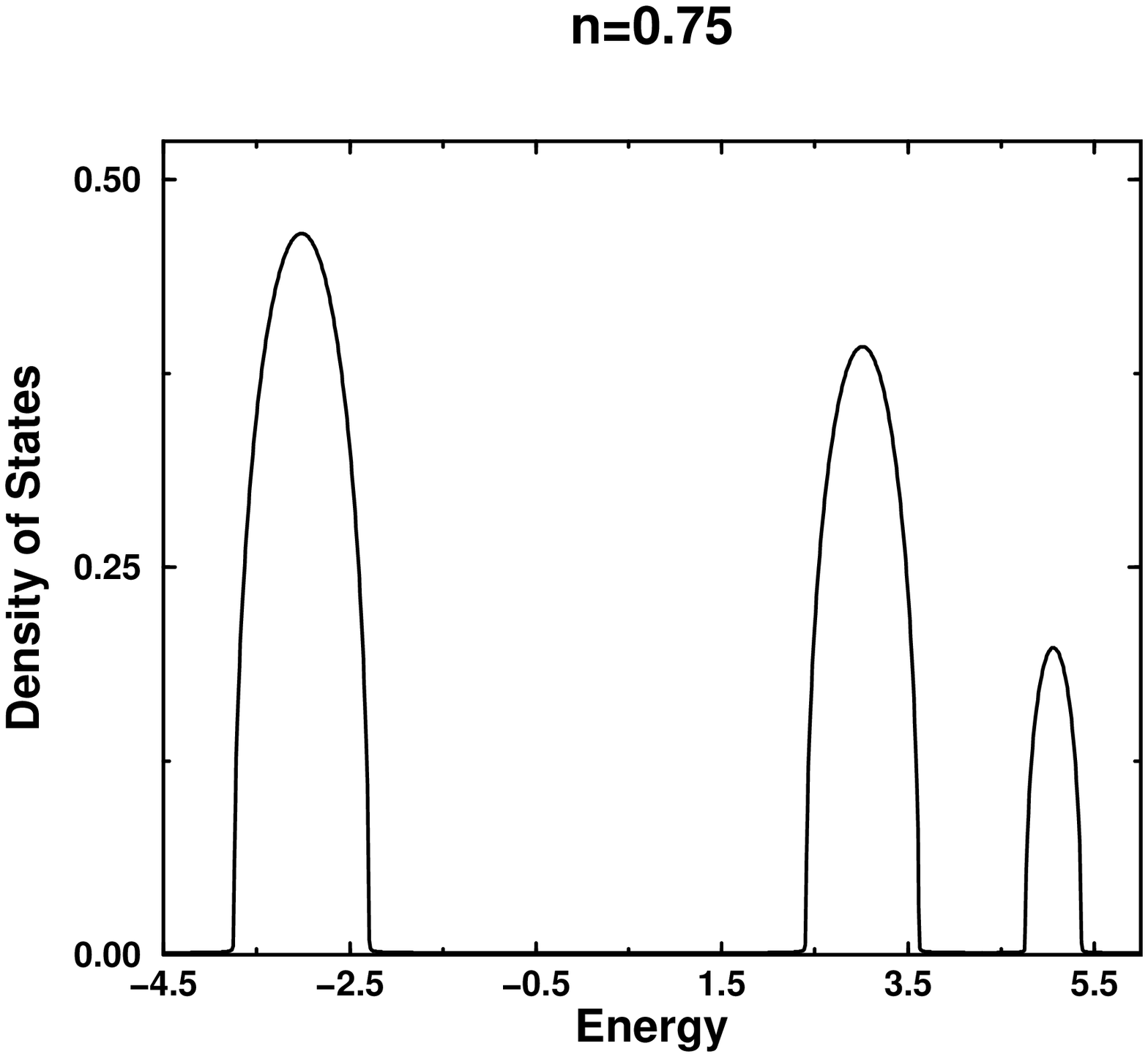}
\epsfxsize=0.5\textwidth\epsffile{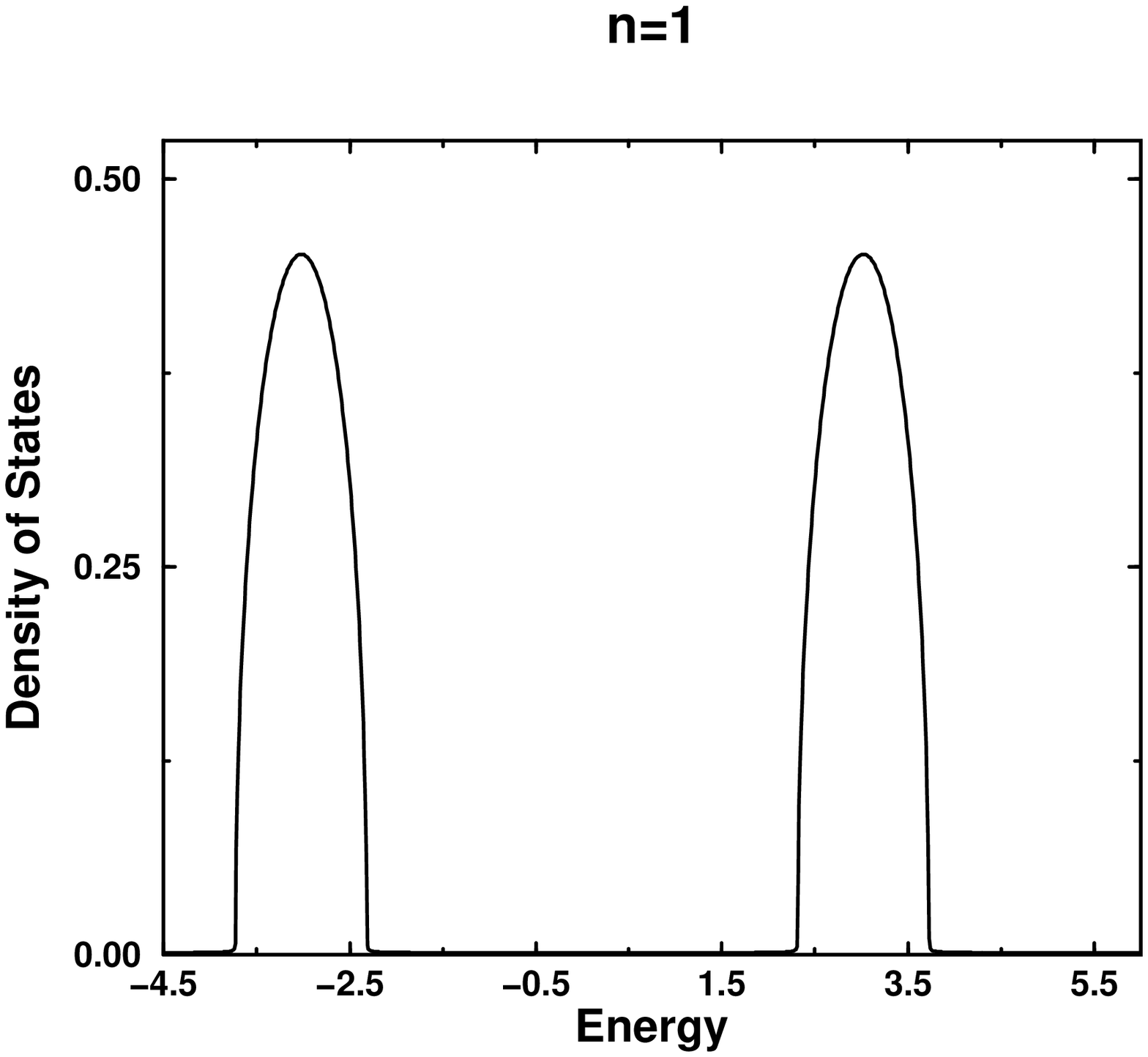}}
\end{center}
\caption{The DOS in the paramagnetic state
for $S=3/2$, $J=4W$, and $n=0,\ 0.25,\ 0.75$ and $1$.
Energy units of $W$ are used.
\label{fpdos}}
\end{figure}

We now consider the zero field paramagnetic state. It may be shown that
$\las {\S}\cdot\bs\ras\rightarrow nS/2$ as $J\rightarrow\infty$, and
$\las {\S}\cdot\bs\ras$ will be very near to this limit as long as
$JS\stackrel{\tiny >}{\tiny _\sim}2W$. We make this approximation in
\fref{fpdos}, in which the paramagnetic state DOS is plotted
for $S=3/2$ and $J=4W$ for various $n$; this has the effect of removing the
weak band centred on $-J(S+1)/2$. It may be seen that as $n$ increases from
0 the band near
$J(S+1)/2$ is reduced in weight and a new band appears near $JS/2$, until
at $n=1$ no weight remains in the band near $J(S+1)/2$. The weight in the
band
near $-JS/2$ is $(S+1-n/2)/(2S+1)$ per spin, so if $JS$ is sufficiently
large
to separate the bands ($JS\stackrel{\tiny >}{\tiny _\sim}2W$) this
band will just be filled at $n=1$ producing a Mott insulator, as discussed 
in
\sref{sAL}. 

We can understand \fref{fpdos} by expanding \eref{eparaGF} in 
partial fractions,
\begin{equation}
\fl G(\en)=\frac{1}{2S+1}\sum_{\alpha=\pm}
\left(\frac
{(S+1)(n/2)^{\alpha}+\las {\S}\cdot\bs\ras}
{E(\en)
-\alpha JS/2}+\frac
{S(n/2)^{\alpha}-\las {\S}\cdot\bs\ras}
{E(\en)
+\alpha J(S+1)/2}\right).\label{eplocal1}
\end{equation}
Comparing this equation with
\eref{edwardsal} in the paramagnetic state we see that
$G(\en)=g(E(\en))$, and since $E(\en)\rightarrow\en$ as $t_{ij}\rightarrow 0$
it is clear how the bands centred on the peaks of the atomic limit spectral
function arise as $t_{ij}$
is switched on.

In the strong-coupling limit $J\rightarrow\infty$, which is
taken with the energy origin shifted by $-JS/2$, \eref{eplocal1} simplifies
 to
\begin{equation}
G(\en)=\frac{(S+1-n/2)/(2S+1)}{E(\en)},
\end{equation}
which corresponds to an elliptical band of weight $(S+1-n/2)/(2S+1)$ and
bandwidth $2W\sqrt{(S+1-n/2)/(2S+1)}$. The band-narrowing factor is
expected to
depend upon the form of the bare DOS, as pointed out by Kubo \cite{rKubo2},
but it is interesting to note that ours is the square root of 
the
factor obtained by Brunton and Edwards \cite{rBruntonEdwards} using a 
different
method.

\section{Comparison with the hole CPA}\label{shole}

In \cite{rKubo1} Kubo and Ohata used a canonical transformation to derive
an
effective Hamiltonian $H'$ for the $J\rightarrow\infty$ limit of the DE
model.
They then mapped $H'$ onto an effective Hamiltonian $H''$ for holes, in
which
conduction holes hop with spins aligned antiparallel to local spins
$(S+1/2)$.
In \cite{rKubo2} Kubo studied $H''$ for $n=1$, using a dynamical CPA
similar to his $n=0$ CPA mentioned above. There is no obvious reason for
one
of these CPA's to be a better approximation than the other, so we would
like
our many-body CPA to agree with both of Kubo's approximations
in the appropriate limits.
In this section we set $n=1$ and $J=\infty$ in our $S=1/2$ Green function
equation \eref{egfs} and compare the result with Kubo's hole CPA.

For $S=1/2$ Kubo's $n=1$ Green function is given by
\begin{equation}
\label{ekubon1}
G_{\u}=\frac{3/2\,({\tilde E}_{\dd}+{\tilde E}_{\u})
+(2{\tilde E}_{\dd}+{\tilde E}_{\u})\las S^z+\s^z\ras+
2({\tilde E}_{\dd}-{\tilde E}_{\u})\las S^z\s^z\ras}
{2/3\,(2{\tilde E}_{\dd}+{\tilde E}_{\u})(2{\tilde E}_{\u}+
{\tilde E}_{\dd})}
\end{equation}
where ${\tilde E}_{\s}(\en)=\en-3W^2/8\,G_{\s}(\en)$. If we set $n=1$ and
$J=\infty$ in \eref{egfs} we obtain
\begin{equation}
G_{\u}= \frac{(E_{\dd}+3E_{\u})/4 + (E_{\dd}+E_{\u})\las S^z+\s^z\ras/2
+(E_{\dd}-E_{\u})\las S^z\s^z\ras}
{E_{\u}(E_{\u}+E_{\dd})}.
\label{eusn1}
\end{equation}
For $n=1$ and $J=\infty$ the relation $\las S^z\ras=\las \s^z\ras$ holds,
and we have used this to write both expectations as $\las S^z+\s^z\ras/2$.
Unfortunately the Green functions defined by \eref{ekubon1} and
\eref{eusn1}
are not equal; this may be
seen by considering the zero-field paramagnetic state where they reduce to
\numparts
\begin{eqnarray}
G(\en)=\frac{1/2}{\tilde E(\en)}=\frac{1/2}{\en-3W^2/8\,G(\en)}\\
G(\en)=\frac{1/2}{E(\en)}=\frac{1/2}{\en-W^2/4\,G(\en)}
\end{eqnarray}
\endnumparts
respectively, and these equations correspond to bands with different widths.
This appears to be a limitation of our CPA, which we would ideally like
to interpolate continuously between Kubo's CPA's. A possible
explanation for this disagreement is our neglect of resonance
broadening corrections in \sref{sCPA}.

We could attempt to include resonance broadening corrections in our
approximation by writing more equations of motion for these terms. It is
however
difficult to find approximations that give a closed set of equations
for the new Green functions introduced without
spoiling the self-consistency of our approximation. 
An alternative is to use an interpolation scheme containing
arbitrary parameters which are chosen to yield
as many correct moments of the spectral function
as possible \cite{rNolting}.

We first note that an empirical modification of \eref{enew}, of the form
\numparts
\begin{eqnarray}\label{e42a}
E_{\up}(\en)= \en-\frac{W^2}{4}\left[G_{\up}(\en)+
\frac{1}{2}G_{\down}(\en)\right]\\
E_{\down}(\en)= \en-\frac{3 W^2}{8}G_{\down}(\en),\label{e42b}
\end{eqnarray}
\endnumparts
maps \eref{eusn1} for $G_{\up}$ onto Kubo's equation \eref{ekubon1}.
This modification is of the same type as Hubbard's resonance broadening
correction to the strong coupling ($U\rightarrow\infty$) limit of his model.
This suggests that we should introduce an interpolation formula for $E_{\s}$
which reduces to \eref{e42a} and \eref{e42b} in the limits $n=0$ and $n=1$,
respectively. Unfortunately so far work along these lines has not proved
successful.

\section{Resistivity formula}\label{sRho}

We now derive a formula for the DC resistivity $\rho$ of the
paramagnetic
state of our model, taking into account the cubic symmetry of the crystal
and
the local nature of our approximation. The Kubo formula \cite{rRKubo}
states that for a small electric field uniform in space but
oscillatory in time with frequency $\omega$, $\vec{E}({\bf r},t)=\vec{E}_0
\exp(-\i\omega t)$, the conductivity tensor
$\sigma_{\mu\nu}(\omega)$ is given in terms of the current-current
correlation
function by
\begin{equation}
\sigma_{\mu\nu}(\omega)=\frac{\i n\ec^2}{\Omega m\omega}\delta_{\mu\nu}
+\frac{\i N\Omega}{\omega}\lla J_{\mu} ; J_{\nu}\rra_{\omega}
\label{econd0}
\end{equation}
where $\hbar=1$, $\Omega$ is the unit cell volume,
$m$ and $-\ec$ are the
electron mass and charge respectively, and the retarded Green function is
used. $\vec{J}$ is the electric current density operator defined for a
homogeneous system by
\begin{equation}
\vec{J} = -\frac{\ec}{N\Omega}\sum_{\k\s}{\vec{v}_{\k}}n_{\k\s}
\label{eJdef}
\end{equation}
where the velocity $\vec{v}_{\k}=\nabla_{\k}t_{\k}$.
In our case the conductivity is a real scalar so from \eref{econd0}
\begin{equation}
\s=-N\Omega\,\lim_{\omega\rightarrow 0}{\rm Im}
\left[\frac{\lla J_x;J_x\rra_{\omega}}{\omega}\right].
\label{econd1}
\end{equation}
Now from \eref{eJdef}
\begin{equation}
\lla J_x;J_x\rra_{\omega}=\frac{\ec^2}{3(N\Omega)^2}
\sum_{\k\k'\s\s'}{\vec{v}_{\k}}\cdot{\vec{v}_{\k'}}
\lla n_{\k\s};n_{\k'\s'}\rra_{\omega},
\label{eJ0}
\end{equation}
so we need an approximation to the
two-electron Green function
$\llas n_{\k\s};n_{\k'\s'}\rras_{\omega}$.

Since the self-energy is independent of momentum
a reasonable approximation is to assume that the irreducible
vertex function is also independent of momentum. In infinite dimensions
where the self-energy is rigorously local the
momentum-dependent contribution
of the irreducible vertex function vanishes.
In this case
the contribution of the vertex correction to $\s$ vanishes owing to the
different parities of $\vec{v}_{\k}$ and $t_{\k}$ in $\k$, and we can
evaluate
the two-electron Green function
$\llas n_{\k\s};n_{\k\s}\rras_{\i\nu}$ in the bubble approximation, obtaining
\numparts
\begin{eqnarray}
\fl
\frac{1}{\beta}\sum_{m}G_{\k\s}(\i\omega_m+\mu)G_{\k\s}(\i\omega_m+\i\nu+
\mu)
\\
 \lo= \frac{1}{\beta}\int{\rm d}\en\int{\rm d}\eta\sum_m
\frac{A_{\k\s}(\en)A_{\k\s}(\eta)}{\left(\i\omega_m+\mu-\en\right)
\left(\i\omega_m+\i\nu+\mu-\eta\right)}
\end{eqnarray}
\endnumparts
where the $\i\omega_m$'s are odd Matsubara frequencies and $\i\nu$ is an
even
Matsubara frequency and the spectral representation of the one-electron
Green
function, $A_{\k\s}(\en)=-{\rm Im}\,G_{\k\s}(\en)/\pi$, has been used.
The sum over $m$ can be evaluated using Cauchy's residue theorem, and
following
analytic continuation to the real axis and a shift in $\eta$ we obtain
\begin{equation} 
\fl \lla n_{\k\s};n_{\k\s}\rra_{\omega}=\int{\rm d}\en\int{\rm d}\eta
\frac{A_{\k\s}(\en)A_{\k\s}(\eta+\en)}{\omega-\eta}\left[f(\en-\mu)-
f(\en+\eta-\mu)\right].
\label{efermi}
\end{equation}

In the paramagnetic state $G$ is $T$-independent if we assume
$\las {\S}\cdot\bs\ras=nS/2$.
The Fermi functions in \eref{efermi} give the conductivity $\s$ a weak
$T$-dependence but we neglect this dependence and calculate at $T=0$,
considering our calculation to apply to the $T> T_{\rm C}$ state
however. Equations \eref{econd1}, \eref{eJ0}, and  \eref{efermi} then imply
that
\begin{equation}
\s\approx \frac{2\pi \ec^2}{3N\Omega}\sum_{\k}\vec{v}_{\k}^2 A_{\k}(\mu)^2.
\label{econd2}
\end{equation}

Now for the simple cubic band
$t_{\k}=-2t[\cos(k_x a)+\cos(k_y a)+\cos(k_z a)]$, where
$a$ is the lattice constant, and \eref{econd2} can be simplified using 
Gauss' theorem:
\begin{equation}
\s=\frac{2\pi \ec^2}{3Na}\sum_{\k} t_{\k}\phi(t_{\k})
\label{econd3}
\end{equation}
where $\phi'(t_{\k})=A_{\k}(\mu)^2$. This is a legitimate definition since
$A_{\k}$ depends on $\k$ only through $t_{\k}$ in the local
approximation. We introduce the cubic bare DOS $D_{\rm c}(\en)$,
and \eref{econd3} becomes
\begin{equation}
\sigma=\frac{2\pi\ec^2}{3 a}\int{\rm d}\en\,\en\,D_{\rm c} (\en)\,
\phi(\en).
\label{econd4}
\end{equation}
Integrating $A_{\en}(\mu)^2$ with respect to $\en$
it may be shown that
\begin{equation}
\fl \phi(\en)=\frac{1}{2\pi^2}\left[\frac{\en+\Sigma'(\mu)-\mu}
{(\Sigma''(\mu))^2+(\en+\Sigma'(\mu)-\mu)^2}+\frac{1}{\Sigma''(\mu)}
\tan^{-1}\left(\frac{\en+\Sigma'(\mu)-\mu}{\Sigma''(\mu)}\right)\right]
\label{ephi}
\end{equation}
where $\Sigma'(\mu)$ and $\Sigma''(\mu)$ are the real and imaginary parts
of $\Sigma(\mu)$ respectively.

If $D_{\rm c}(\en)$ is replaced with the Gaussian $D_{\rm g}(\en)$
corresponding to an infinite dimensional approximation \cite{rMuller},
\eref{econd4} may be simplified by integrating by parts:
\begin{equation}
\sigma=\frac{\pi\ec^2 W^2}{9a}\int\d\en\, D_{\rm g}(\en)A_{\en}(\mu)^2.
\label{efsigma}
\end{equation}
Note that this conductivity formula is of the same form as Furukawa's
\cite{rFurukawa}.

\section{Resistivity calculations}\label{sResults}

In this section
we will use formulas \eref{econd4}, \eref{ephi}, and \eref{efsigma}
to calculate $\rho=\s^{-1}$,
making simple analytic approximations for $D_{\rm c}(\en)$.
Note that in SI units \eref{econd4} and \eref{efsigma} must
be divided by $\hbar$.

\subsection{Elliptical DOS}

\begin{figure}[htbc]
\begin{center}
\leavevmode
\hbox{%
\epsfxsize=0.5\textwidth \epsffile{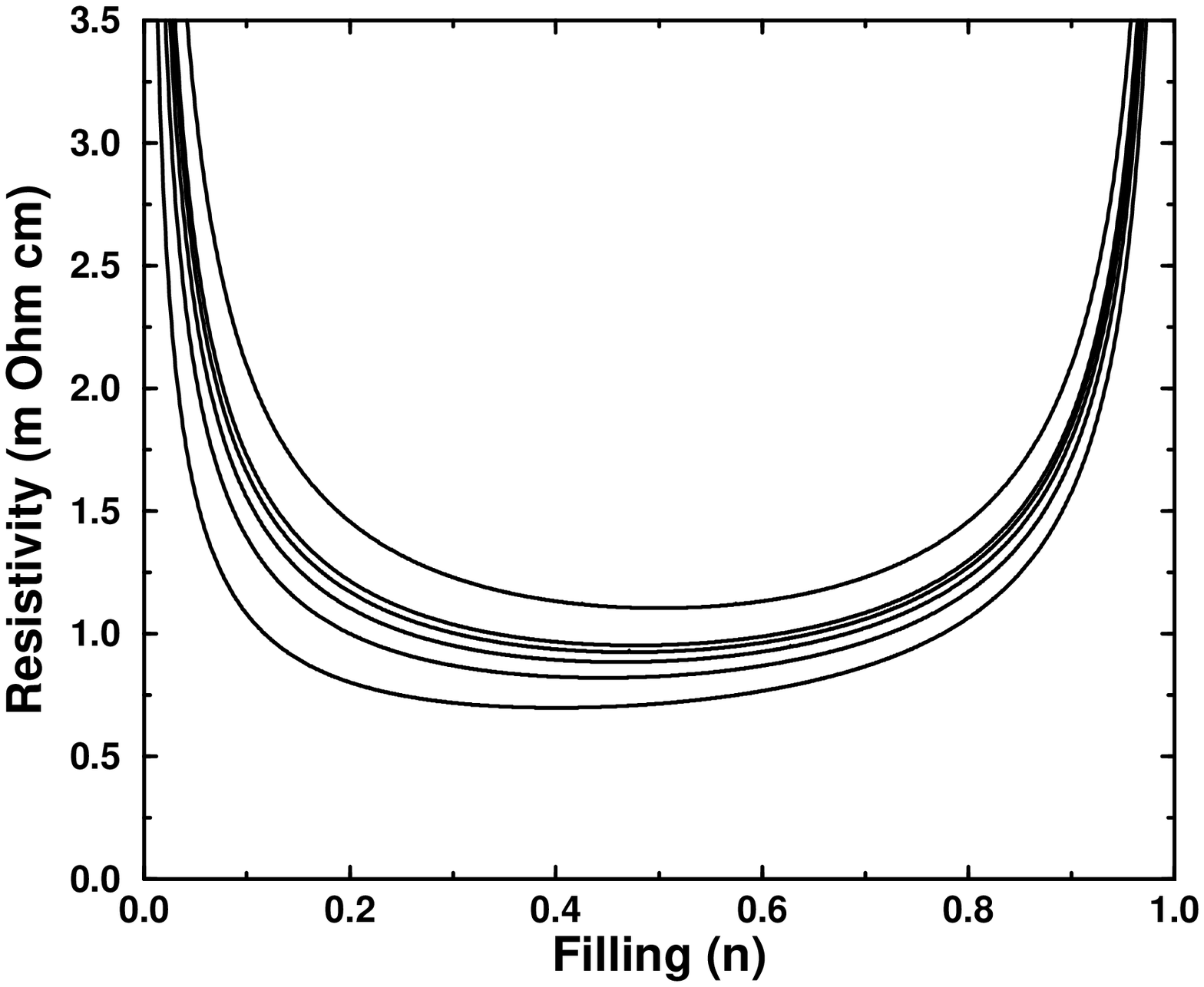}}
\end{center}
\caption{The zero field paramagnetic state resistivity $\rho$
(in m$\Omega$cm) 
versus filling $n$, plotted for $J=\infty$, $a=5$\AA, and $S=$1/2, 1, 3/2,
 2,
5/2, and $\infty$, $\rho$ increasing with $S$.
The elliptical DOS and resistivity formula \eref{econd4} are used.
\label{res_fig1}}
\end{figure}

We first calculate $\rho$ for the elliptical DOS $D_{\em e}(\en)$.
We take the strong-coupling limit
$J\rightarrow\infty$ for simplicity; the precise value of $J$ is
unimportant in
the strong-coupling regime $JS\stackrel{\tiny >}{\tiny _\sim}2W$.
 $\rho$ is
plotted against $n$ for various $S$ in \fref{res_fig1}.
It may be seen
that $\rho$ correctly diverges at $n=0$ and $n=1$, but the size of
$\rho$ and $\partial\rho/\partial n$ for $n\sim 0.8$ are much too small to
explain experiment, at least in some materials. For example
Urushibara \etal \cite{rTokura}, who measured $\rho$ for
La$_{n}$Sr$_{1-n}$MnO$_3$, found (for $T$ well above $T_{\rm C}$) that
$\rho\sim$ 20 m$\Omega$cm for $n\sim 0.8$ and that $\rho$ drops by
about an order of magnitude as $n$ is reduced from 0.85 to 0.7.
We therefore agree with Millis \etal \cite{rMillis}
that the bare DE model does not explain the paramagnetic state of the CMR
materials. We also
find that the $T$-dependence of $\rho$ is always of the metallic
$\partial\rho/\partial T>0$ form in our approximation, whereas Urushibara
\etal
find a crossover between metallic behaviour at $n=0.7$ to insulating
behaviour
at $n=0.85$. It should be noted that $\rho$ is independent of $W$ here, and
the size of $J$ and $S$ and the choice of resistivity formula 
\eref{econd4} or \eref{efsigma}, using $D_e$ in both cases,
does not have a large effect.

\subsection{Lorentzian DOS}

In order to make contact with Furukawa's work \cite{rFurukawa} we now
consider
the effects of approximating the true cubic DOS $D_{\rm c}(\en)$ with the
Lorentzian $D_{\rm l}(\en)$,
which is somewhat less realistic than the elliptical
DOS approximation owing to its slowly decaying tails.
Our equations are very simple in this case: it may be shown that
$J(\en)=-iW$ and our equations for the Green function become explicit; 
switching on $W$ here merely broadens the peaks of the atomic limit
spectral function into Lorentzians with the same width parameter $W$ as the
bare DOS. Our approximation and Furukawa's become
very similar in this case;
in the classical $S\rightarrow\infty$ limit for example our local
Green function is equal to Furukawa's.

The slowly decaying form of the Lorentzian means that the precise value
of $J$ has more effect here than in the case of the elliptical DOS: the
approximation $\las {\S}\cdot\bs\ras\approx nS/2$ only
holds accurately for very large $J$ and we only obtain a true insulator at
half-filling for $J=\infty$ for instance. For simplicity however we take
the
limit $J\rightarrow\infty$ (with energy origin shifted by $-JS/2$)
rather than use the finite value used by Furukawa.
In the paramagnetic state the local Green function $G$ is then
\begin{equation} 
G(\en)=\frac{S+1-n/2}{2S+1}\frac{1}{\en+iW},
\end{equation}
and the self-energy and chemical potential are
\numparts
\begin{eqnarray}
\Sigma(\en)= -\left(\frac{2S+n}{2(S+1)-n}\right)(\en+iW)\\
\mu= W\tan\left[\frac{\pi}{2}\left(\frac{(4S+3)n-2(S+1)}{2(S+1)-n}\right)
\right].
\end{eqnarray}
\endnumparts

If we use our finite-dimensional
formula \eref{econd4} to calculate $\rho$ with the
Lorentzian DOS we must use a
mixed approximation to the DOS, calculating $\phi$ using the Lorentzian
approximation but using the elliptic approximation for $D_{\rm c}(\en)$,
in order
for the integral in \eref{econd4} to converge. This problem does not
arise when using the infinite dimensional conductivity formula
\eref{efsigma} so $D_{\rm l}$ can be used throughout.
The choice of conductivity formula \eref{econd4} (with mixed approximation)
or \eref{efsigma}
does not qualitatively alter the results, but since the calculation using
\eref{econd4} has already been published \cite{rThey} we use \eref{efsigma}
here. It also seems more satisfactory to use the same approximation for
the DOS throughout. We believe conductivity formula \eref{econd4}
to be more realistic however, 
as it does not rely on the infinite dimensions approximation.

\begin{figure}[htbc]
\begin{center}
\leavevmode
\hbox{%
\epsfxsize=0.5\textwidth \epsffile{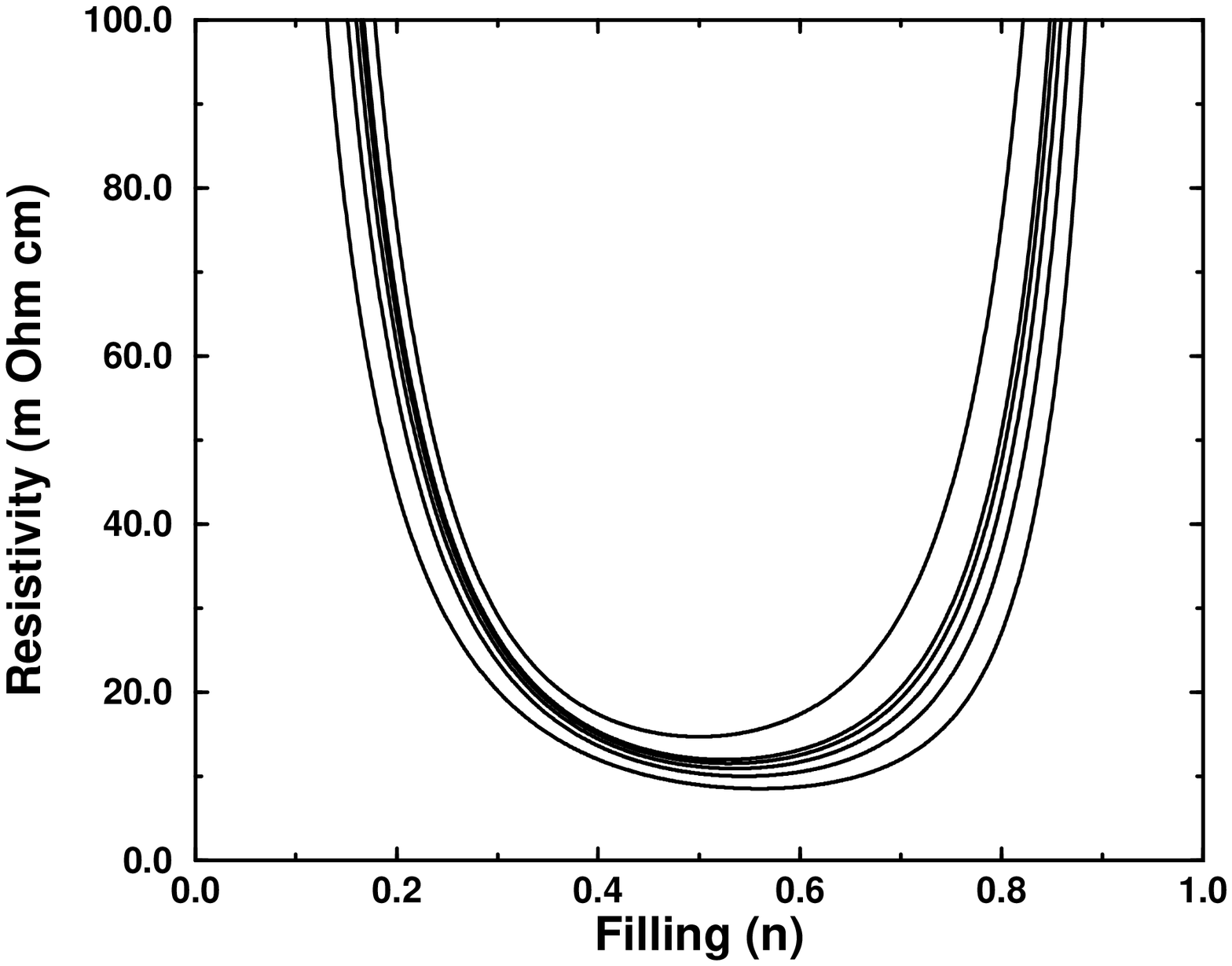}}
\end{center}
\caption{The zero field
paramagnetic state resistivity $\rho$ (in m$\Omega$cm) versus
filling $n$, plotted for $J=\infty$, $a=5$\AA, and $S=$1/2, 1, 3/2, 2, 5/2,
and $\infty$, $\rho$ increasing with $S$. The Lorentzian DOS and
resistivity
formula \eref{efsigma} are used.
\label{res_fig2}}
\end{figure}

$\rho$ is plotted against $n$ for various $S$
in \fref{res_fig2}.
Note that the result is independent of $W$.
 As before we obtain insulating
behaviour at $n=0$ and $n=1$, but now 
$\rho$ and $\partial\rho/\partial n$ are of the same order of magnitude as
experiment for $n\sim 0.8$, and our results are also consistent with those 
of
Furukawa. Even better agreement can be obtained by reducing $J$, which has 
the
effect of reducing $\rho$ and $\partial\rho/\partial n$ for $n\sim 0.8$.
Furukawa's result--- that the DE model's prediction for
$\rho$ is of the same order of magnitude as the experimental value---
is therefore due to the unphysical approximation he used for the cubic DOS.

\section{Summary}\label{sSummary}

In addition to theoretical predictions of the properties of
the DE model, experiments \cite{rZhao} also suggest that coupling to
lattice degrees of freedom is a necessary ingredient in a model of CMR
systems.
However, we take the view that a good understanding of the bare DE model is
necessary before more accurate and complicated models can be tackled.
Accordingly in this paper we have presented a many-body CPA for the DE
model
and a compatible formula for the resistivity $\rho$; our main
result is the reconciliation of the calculations of Millis \etal
\cite{rMillis} and Furukawa \cite{rFurukawa} of $\rho$ in the paramagnetic
state. We have confirmed that single-site scattering within the DE model is
inadequate to describe the resistivity of CMR systems.

Now the alloy CPA is known to give good results for a wide range of systems
 in 
which localization effects are unimportant. This should be the case here
since,
although some authors have claimed otherwise \cite{rVarma}, localized
states
are expected to occur for the bare DE model only in the tails of the band; 
we therefore expect our formulae for $G_{\s}$ to be good approximations.
Nevertheless several problems exist with our approximation. Firstly, in
common
with the usual alloy CPA the imaginary part of the self-energy does not
vanish
at the Fermi level at $T=0$ so we do not obtain a true Fermi liquid; this
should not be important in the high temperature paramagnetic regime that we
have considered but the problem could perhaps be fixed following the
approach
of Edwards and Hertz to the Hubbard model \cite{rEdwardsHertz}. Secondly,
we have shown that for finite $S$ the saturated ferromagnetic state is
never
consistent in our approximation. Real CMR systems are believed to exhibit
strong ferromagnetism in some regimes, and it is generally expected that
the DE
model has regions of strong ferromagnetism. Finally,
although we recover Kubo's $n=0$ CPA in the
low-density limit, we do not recover his strong-coupling $n=1$ CPA
\cite{rKubo2} at half-filling. This appears to be a shortcoming of our
approximation, possibly due to our neglect of resonance broadening
corrections.
These problems and possible remedies will be discussed in a later paper
 in
which we will concentrate on the magnetic susceptibility of our CPA.

\ack

KK was supported by EPSRC grant number GR/L90804 and JSPS grant number 09640453
and ACMG by an EPSRC studentship and a Monbusho REFYFR grant.


\section*{References}
\begin{thebibliography}{99}
\def\paper#1#2#3#4#5{#1 19#5 #2 {\bf #3} #4}
\def\abook#1#2#3#4{#1 19#4 {\it #2} (#3)}

\bibitem{rRamirez}
\paper{Ramirez AP}{\JPCM}{9}{8171--8199}{97}
\bibitem{rZener}
\paper{Zener C}{\PR}{82}{403--405}{51}
\bibitem{rFurukawa}
\paper{Furukawa N}{\JPSJ}{63}{3214--3217}{94}
\bibitem{rMillis}
\paper{Millis AJ, Littlewood PB and Shraiman BI}{\PRL}{74}{5144--5147}{95}
\bibitem{rMillis2}
\paper{Millis AJ, Shraiman BI and Mueller R}{\PRL}{77}{175--178}{96}
\bibitem{rZhao}
\paper{Zhao G, Conder K, Keller H and M\"uller KA}{Nature}{381}{676--678}
{96}
\bibitem{rUs}
\paper{Edwards DM, Green ACM and Kubo K}{{\it Physica} B, In Press}{}{}{98}
\bibitem{rThey}
\paper{Kubo K, Edwards DM, Green ACM, Momoi T and Sakamoto H}
{In Press}{}{}{98}
\bibitem{rMe}
\abook{Green ACM}{Correlated electrons in heavy fermion and double exchange
systems}{London: Imperial College PhD thesis}{98}
\bibitem{rHubbard3}
\paper{Hubbard J}{\PRS}{281}{401--419}{64}
\bibitem{rFukuyama?}
\paper{Fukuyama H and Ehrenreich H}{\PR B}{7}{3266}{73}
\bibitem{rKubo2}
\paper{Kubo K}{\JPSJ}{33}{929--935}{72}
\bibitem{rRKubo}
\paper{Kubo R}{\JPSJ}{12}{570--586}{57}
\bibitem{rKubo}
\paper{Kubo K}{\JPSJ}{36}{32--38}{74}
\bibitem{rMuller}
\paper{M\"uller-Hartmann E}{\ZP B}{74}{507--512}{89}
\bibitem{rBruntonEdwards}
\paper{Brunton RE and Edwards DM}{\JPCM}{10}{5421--5431}{98}
\bibitem{rKubo1}
\paper{Kubo K and Ohata N}{\JPSJ}{33}{21--32}{72}
\bibitem{rNolting}
\paper{Herrmann T and Nolting W}{\PR B}{53}{10579--10588}{96}
\bibitem{rTokura}
\paper{Urushibara A, Moritomo Y, Arima T, Asamitsu A, Kido G and
Tokura Y}{\PR B}{51}{14103--14109}{95}
\bibitem{rVarma}
\paper{Varma CM}{\PR B}{54}{7328--7333}{96}
\bibitem{rEdwardsHertz}
\paper{Edwards DM}{\JPCM}{5}{161--170}{93}




\end{thebibliography}
\end{document}